\documentclass[prd,nofootinbib,aps,superscriptaddress,tightenlines,preprintnumbers]{revtex4} 
\begin{document}

\title{THE LEE WICK STANDARD  MODEL}

\author{MARK B. WISE$^*$}

\address{Division of Physics Mathematics and Astronomy, California Institute of Technology,\\
Pasadena, California  91125, USA\\
$^*$E-mail: wise@theory.caltech.edu\\}

\begin{abstract}
This article reviews some recent work on a version of the standard model (the Lee-Wick standard model) that contains higher derivative kinetic terms that improve the convergence of loop diagrams removing the quadratic divergence in the Higgs boson mass. Naively higher derivative theories of this type are not acceptable  since  the higher derivative terms either cause instabilities (from negative energies) or a loss of unitarity (from negative norm states).  Lee and Wick provided an interpretation for such  theories arguing that theories with  higher derivative kinetic terms can be unitary and stable if the states associated with the massive  propagator poles, that arise from the higher derivatives,  have widths and hence decay and are not in the spectrum of the theory.  
\end{abstract}

\keywords{higher derivative; Lee-Wick}
\maketitle
%\bodymatter

\section {Introduction}\label{aba:sec1}

Its a pleasure to attend this meeting to help celebrate Misha Shifman's 60'th birthday. I have spent much of my career following up on the ideas of Misha and his collaborators and am grateful for the many things I have learned from studying their papers. I have fond memories, from my graduate student days,  of waiting anxiously in the SLAC library on Friday afternoon for the new preprints to be put out and hoping there would be something from the ITEP group appearing that week. My recollection is that the paper used was roughly the same quality as toilet paper, the equations were  hand written, but the physics was fantastic.

I am going to talk today about some very speculative work I have been doing over the last year or so with Benjamin Grinstein, Donal O'Connell and other collaborators. We have been studying theories with higher derivative kinetic terms. In this work the higher derivatives kinetic terms are not treated as a perturbation but re-summed into the denominator of the propogator. This improves the ultraviolet convergence of Feynman diagrams but leads to a host of potential problems. At first glance, it seems that such theories must either be unstable or not unitary.  However, Lee and Wick (LW) argued that this is not necessarily the case~\cite{LW}. In the next section I  review the ideas of Lee and Wick using  a simple toy model. Then I discuss the extension to the  standard model (the LW standard model ~\cite{GOW}) and explain why the resulting theory has no quadratic divergences in the Higgs mass and hence no hierarchy puzzle. Concluding remarks are made in the final section. There is another interpretation for theories with higher derivative kinetic terms that I will not discuss in this lecture~\cite{bender}.

Another case where ultraviolet divergences cause a problem is quantum gravity. It may be that the ideas of Lee and Wick can be applied to get a sensible quantum theory of gravity~\cite{tomboulis}. The present best candidate for a quantum theory of gravity is String Theory, and most theoretical physicists believe it is the correct theory. However,  there is no experimental evidence to support the conjecture that String Theory is realized in nature. I will not say anything about Lee Wick theories of gravity in this lecture taking the attitude that it is best to start off trying to understand Lee Wick  versions of ordinary gauge field theories before addressing gravity.

In this talk I will try to explain some of what is known about  LW- theories and the LW standard model in particular. There is  some evidence a Lorentz invariant unitary $S$-matrix can be defined perturbatively, however,  these theories have not been extensively studied and more work is needed to firmly establish that they are  consistent physical theories. A very good introduction to the ideas of Lee and Wick can be found in lectures by Sidney Coleman~\cite{coleman}.

\section{A Toy Model}

Consider a model with a single real scalar field $\hat \phi$ and Lagrange density,
\begin{equation}
\label{lagrangehd}
{\cal L}_{\rm hd}={\partial}_{\mu}{\hat \phi}{\partial}^{\mu}{\hat \phi}-{1 \over 2 M^2}\left(\partial^2 {\hat \phi}\right)^2-{1 \over 2}m^2{\hat \phi}^2-{1 \over 3!}g{\hat \phi}^3.
\end{equation}
This Lagrange density gives the ${\hat \phi}$ propogator,
\begin{equation}
{D}_{\hat \phi}(p)={i \over p^2-p^4/M^2-m^2}.
\end{equation}
For $M\gg m$ the above propagator has poles at $p^2 \simeq m^2$ and also at $p^2 \simeq M^2$. The residue of the pole at $p^2=M^2$ has the wrong sign which, naively,  signals an instability or a negative norm state that causes a lack of unitarity.  

Imagine, for the moment, doing classical physics with the higher derivative Lagrange density in Eq.~(\ref{lagrangehd}) by treating the interaction term proportional to $g$ using perturbation theory. Because of the higher derivative kinetic term specifying the usual initial conditions is not enough to get a unique solution. In addition a  future boundary condition, that removes the  classically unstable solutions, is imposed.  

Before discussing the quantum mechanical interpretation of this theory that Lee and Wick gave it is useful to provide another formulation where an auxillary field is introduced and the higher derivative term is absent.  The field theory with two real scalars $\hat \phi$ and $\tilde \phi$ with Lagrange density
\begin{equation}
\label{renormlagrange}
{\cal L}={1 \over 2} \partial_{\mu}{\hat \phi} \partial ^{\mu}{\hat \phi}-{1 \over 2}m^2 {\hat \phi}^2
-{\tilde \phi} \partial^2 {\hat \phi}+{1 \over 2}M^2{\tilde \phi}^2- {1 \over 3}g {\hat \phi}^3,
\end{equation}
is equivalent to the field theory with the Lagrange density in Eq.~(\ref{lagrangehd}). To see this note that  the LW scalar field $\tilde \phi$ occurs in quadratic and linear terms in Eq.~(\ref{renormlagrange}). Hence the functional integral over $\tilde \phi$ can be done exactly and this reproduces Eq.~(\ref{lagrangehd}). Equivalently the field $\tilde \phi$ can be removed from Eq.~(\ref{renormlagrange}) using its equations of motion and this also reproduces Eq.~(\ref{lagrangehd}).

The interpretation of the fields $\hat \phi$ and $\tilde \phi$ in Eq.~(\ref{renormlagrange}) is obscured by the fact that the kinetic terms are not diagonal. To diagonalize them we define,
${\hat \phi}=\phi-{\tilde \phi }$. In terms of the ordinary scalar $\phi$ and LW scalar ${\tilde \phi}$ the Lagrange density in Eq.~(\ref{renormlagrange}) becomes,
\begin{equation}
\label{renormlagrange1}
{\cal L}={1 \over 2} \partial_{\mu}{\phi} \partial ^{\mu}{\phi}-{1 \over 2} \partial_{\mu}{\tilde \phi} \partial ^{\mu}{\tilde \phi}+{1 \over 2}M^2 {\tilde \phi}^2-{ 1\over 2}m^2(\phi - {\tilde \phi})^2
 -{1 \over 3}g( \phi-{ \tilde \phi})^3.
\end{equation}
The LW resonance that is destroyed by the field ${\tilde \phi}$ is unstable since it can decay into two $\phi$ particles.  Its decay width is calculated by doing the standard bubble sum for the $\tilde \phi$ propagator. The self energy $\Sigma$ is the same as for an ordinary particle. The only difference is that each LW scalar propagator comes with an additional minus sign, compared with a normal scalar propagator. Hence,
\begin{eqnarray}
&&{ D}_{\tilde \phi}(p)={-i \over p^2-M^2}+{-i \over p^2-M^2}\left(-i\Sigma(p^2)\right){-i \over p^2-M^2}+\ldots \nonumber \\
&&={-i \over p^2-M^2+\Sigma(p^2)}.
\end{eqnarray}
Treating the LW scalar propagator  in the narrow width approximation we see that the LW field is associated with a resonance, but its width term enters with the wrong sign, {\it i.e.},
\begin{equation}
{ D}_{\tilde \phi}(p)={-i \over p^2-M^2-iM\Gamma},
\end{equation}
with
\begin{equation}
\Gamma={g^2 \over 32 \pi M}\sqrt{1-{4m^2 \over M^2}}.
\end{equation}

Consider an experiment where two beams of stable particles collide in the center of mass and create an ordinary narrow resonance approximately at rest. The resonance has a probability $\Gamma {\rm d}t$ of decaying  in the positive time interval ${\rm d}t$ into the stable particles that created it (we assume this is the only allowed decay channel) . The decay products are observed by detectors that can measure their momentum and energy. From those measurements we can determine the time that the resonance decayed by following the paths of the detected decay products back to the point they were produced.  For a Lee-Wick resonance the flip in the sign of the width term means that the resonance appears to decay earlier than it was produced, {\it i.e.}, it has a probability $\Gamma {\rm d}t$ of decaying in  the negative time interval $-{\rm d}t$. Hence the theory is usually called acausal. I used the phrase ``appears to decay earlier than it was produced"  because the future boundary condition forces the state vector for the colliding beams  to contain a very small component of the decay products in the far past and this component grows with time~\cite{tonder}. The apparent acausal behavior is a result of the non-locality in time.

It seems that this unusual acausal behavior should lead to paradoxes. Can't you use the decay products to turn off the experiment before the collision occurred? You will get a headache if you think too hard  about this issue. If the $S$-matrix is unitary there will not be any paradoxes in scattering experiments. You prepare initial stable particle states (in the distant past) out come final stable particle states (in the far future).  The probabilities for the possible final states add up to one and, furthermore, since the $S$-matrix is a one-to-one map  there are no paradoxes.  Hence, we should pay special attention to whether the $S$-matrix in LW theories is unitary on the space of incoming and outgoing ordinary stable particles.

Now we go beyond the narrow resonance approximation and compare the LW scalar propagator with the propagator for a scalar field $\chi$ that corresponds to an ordinary resonance. For the ordinary resonance,
\begin{equation}
D_{\chi}(p^2)={i \over \pi}\int_{4m^2}^{\infty}{\rm d}s {\rho(s) \over p^2-s+i\epsilon} \simeq {i \over p^2-M^2+iM\Gamma}.
\end{equation}
There are no poles because the $\chi$ field destroys a resonance which is not in the spectrum of the theory. It is the stable decay products that are in the spectrum. For the LW scalar we need to get, in the narrow resonance approximation, a propagator with the opposite overall sign and pole in the opposite side of the real axis. However, the self energy is the same and so the cut piece is the same as for an ordinary resonance. Hence,
\begin{equation}
D_{\tilde \phi}(p^2)={-i \over p^2-M_c^2} +{-i \over p^2-M_c^{*2}}+{i \over \pi}\int_{4m^2}^{\infty}{\rm d}s {\rho(s) \over p^2-s+i\epsilon} ,
\end{equation}
where $M_c^2=M^2+iM\Gamma$.
To get a unitary theory Lee and Wick advocate treating the poles at $M_c^2$ and $M_c^{*2}$ as resonances that are not in the spectrum of the theory. They correspond to negative norm states and so if they were in the spectrum the theory would not be unitary. However, they are poles in the $\tilde \phi$ propagator so how can we have a unitary $S$-matrix when we exclude the LW particles from the initial and final states?

 To begin with,  consider tree level $\phi \phi$ scattering through a virtual $\tilde \phi$ resonance. The complex poles do not destroy unitarity since the incoming particles have real energies and momentum and so we cannot produce an on shell intermediate state with complex mass. Their contribution to the scattering amplitude is real and it is easy to verify that tree level exchange satisfies the optical theorem. 
 
The real subtleties first occur at one loop in Feynman diagrams that have two LW massive particles in the intermediate state. One can get an intermediate LW two particle on shell state  produced with real incoming momentum and energy . For these diagrams one must provide a definition for loop integrations of the form
 \begin{equation}
 I=\int {{\rm d}^4p \over (2\pi)^4} {-i \over (p+q)^2-M_1^2}{-i \over p^2-M_2^2},
 \end{equation}
where $M_1$ and $M_2$ are the complex LW masses. The $p^0$ integration has four poles. For time like $q^0$ we can go to the frame where ${\bf q}=0$ and the poles are located at, $p^0=\pm \sqrt{{\bf p}^2+M_2^2}$ and $p^0=-q^0 \pm \sqrt{{\bf p}^2+M_1^2}$. At $g=0$ the widths for the LW resonances vanish and the masses $M_1$ and $M_2$ are real. Then the cotour of $p^0$ integration is the usual Feynman contour\footnote{Since the contour can be Wick rotated the scattering amplitudes are Lorentz invariant. Some ways of defining the amplitudes are not Lorentz invariant~\cite{Lorentz}. } . As $g$ increases the contour is defined so that the poles do not cross the contour; a pole which was initially below the contour remains below for example. This leads to a well defined contour that can be Wick rotated unless poles pinch the contour. Pinching can occur because we could have $M_2^2=M_1^{*2}$  and then for some $q^0$ two poles overlap. In this case we define the integral by taking the masses $M_1$ and $M_2$ to be unrelated complex mass parameters so the poles don't overlap. At the end of the calculation $M_2^2$ is set equal to the complex conjugate of $M_1^2$.  One can show that this leads to a unitary Lorentz invariant scattering amplitude.  This prescription is due to Cukowski {\it et. al.}~\cite{CLOP}.  Boulware and Gross discuss the difficulty of implementing it in a path integral formulation~\cite{bg}.

It is not known if all Feynman diagrams in this theory can be treated in a similar way to get a unitary theory. I suspect there is no impediment to doing so. One reason for this is that the $O(N)$ scalar model can be made into a LW theory by adding a higher derivative kinetic term. One can show that the proceedure of Cutkowski {\it et. al. } renders this theory unitary in the large $N$ limit ~\cite{Grinstein:2008bg}, at all orders in perturbation theory. 

It has been argued  that $U(1)$ Lee Wick  gauge theories are Lorentz invariant, unitary and can be defined nonperturbatively~\cite{tonder}. The argument presented by van Tonder is is a little formal and I have not worked through it.

\section{LW-Gauge Bosons and the Hierarchy Puzzle}

In this section we discuss LW theories with non-Abelian gauge bosons. The higher derivative Lagrange density is,
\begin{equation}
{\cal L}_{\rm hd}=-{1 \over 2}{\rm tr}{\hat F}_{\mu \nu}{\hat F}^{\mu \nu}+{1 \over M_A^2}{\rm tr}\left({\hat D}^{\mu}{\hat F}_{\mu \nu}    {\hat D}^{\lambda}{\hat F}_{\lambda} ^{~\nu}    \right),
\end{equation}
where ${\hat F}_{\mu \nu}=\partial_{\mu}{\hat A}_{\nu}-\partial_{\nu}{\hat A}_{\mu}-ig [{\hat A}_{\mu},{\hat A}_{\nu}]$, and ${\hat A}_{\mu}= {\hat A}_{\mu}^{B}T^B$.  Now the higher derivative term contains not only the order $p^4/M_A^2$ contribution to the denominator of the gauge boson propagator, but also interactions. Again we remove the non-renormalizable term by introducing the auxillary  LW field ${\tilde A}_{\mu}$.  Now the Lagrange density becomes,
\begin{equation}
{\cal L}=-{1 \over 2}{\rm tr}{\hat F}_{\mu \nu}{\hat F}^{\mu \nu}-{ M_A^2}{\rm tr}{\tilde A}_{\mu}{\tilde A}^{\mu}+2{\rm tr}{\hat F}_{\mu \nu}{\hat D}^{\mu}{\tilde A}_{\nu},
\end{equation}
where ${\hat D}_{\mu} {\tilde A}_{\nu}=\partial_{\mu}{\tilde A}_{\nu}-ig[{\hat A}_{\mu}.{\tilde A}_{\nu}]$. To diagonalize the kinetic terms we shift the hatted gauge field,
\begin{equation}
{\hat A}_{\nu}={ A}_{\mu}+{\tilde A}_{\nu}.
\end{equation}
In terms of the LW gauge field  ${\tilde A}_{\mu}$ and the ordinary gauge field  ${ A}_{\mu}$ the Lagrange density now becomes,
\begin{eqnarray}
\label{gaugerenorm}
&&{\cal L}=-{1 \over 2}{\rm tr}{ F}_{\mu \nu}{ F}^{\mu \nu}+{1 \over 2}{\rm tr}\left(D_{\mu} {\tilde A}_{\nu}-D_{\nu} {\tilde A}_{\mu}\right)\left(D^{\mu} {\tilde A}^{\mu}-D^{\nu} {\tilde A}^{\mu}\right)\nonumber \\
&&-ig {\rm tr} \left(\left[{\tilde A}_{\mu},{\tilde A}_{\nu}\right] F^{\mu \nu} \right) -{3 \over 2} g^2 {\rm tr}\left(\left[{\tilde A}_{\mu},{\tilde A}_{\nu}\right]\left[{\tilde A}^{\mu},{\tilde A}^{\nu}\right]\right) \nonumber \\
&&-4ig {\rm tr}\left(\left[{\tilde A}_{\mu},{\tilde A}_{\nu}\right]D^{\mu}{\tilde A}^{\nu}    \right)-M_A^2 {\rm tr}\left( {\tilde A}_{\mu}{\tilde A}^{\mu} \right).
\end{eqnarray}

In this formulation the Lagrange density is renormalizable by power counting and the LW gauge field is a vector matter field in the adjoint representation of the gauge group with very specific interactions given by Eq.~(\ref{gaugerenorm}).  

Since there are massive vectors in this theory one should be concerned that the theory will become strongly coupled or violate unitarity because the  scattering of the longitudinal polarizations  grows with energy. After all, in the standard model with no Higgs boson but a mass term for the W-bosons the longitudinal polarization\footnote{We use  $L$ to denote a Longitudinal polarization and $T$ to denote a transverse polarization.} scattering amplitude grows with energy as, ${\cal A}(LL \rightarrow LL)\sim g_2^2E^2/M_W^2$. This bad behavior arises because the components of the longitudinal polarization vector for a vector boson with mass $M$, $\epsilon_L(p)=(|{\bf p}|, E(p){\bf {\hat p}})/M$, grow with energy $E$.  However the LW massive gauge boson has couplings constrained  by an unbroken gauge invariance and so it can be shown that there is no bad growth with energy~\cite{gow2}. In fact, ${\cal A}(LL\rightarrow LL) \sim g^2M_A^2/E^2$, $ {\cal A}(LL \rightarrow LT) \sim g^2 M_A/E$ and ${\cal A}(LL\rightarrow TT)\sim g^2$.  If the higher derivative Lagrange density could not be rewritten in terms of a renormalizable Lagrange density using an auxillary field we would have indeed found that the scattering amplitudes grow with energy. 

We are almost ready to discuss the hierarchy problem in the LW standard model. To do that we add a scalar in the fundamental of the gauge group $\hat \phi$. Its interactions are governed by the Lagrange density,
\begin{equation}
\label{hdscalar}
{\cal L}_{\rm hd}=\left({\hat D}_{\mu} {\hat \phi}\right)^{\dagger}\left( {\hat D}^{\mu}{\hat \phi}\right)-{ 1\over M_{\phi}^2}
\left({\hat D}_{\mu} {\hat D}^{\mu}{\hat \phi}\right)^{\dagger}\left({\hat D}_{\nu} {\hat D}^{\nu}{\hat \phi}\right).
\end{equation}
As in the case of the gauge fields the higher derivative term also contributes to the interactions. However, it does not destroy renormalizability because it can be eliminated by introducing the LW field ${\tilde \phi}$. After shifting ${\hat \phi}=\phi-{\tilde \phi}$ Eq.~(\ref{hdscalar}) becomes
\begin{eqnarray}
&&{\cal L}=\left({D}_{\mu} { \phi}\right)^{\dagger}\left( {D}^{\mu}{ \phi}\right)-\left({D}_{\mu} {\tilde \phi}\right)^{\dagger}\left( {D}^{\mu}{\tilde \phi}\right)+M_{\phi}^2 {\tilde \phi}^{\dagger}{\tilde \phi}   \\
&&+ig\left(D^{\mu} \phi \right)^{\dagger}{\tilde A}_{\mu}^B T^B\phi+g^2{\phi}^{\dagger}\left({\tilde A}_{\mu}^B T^B{\tilde A}_{\mu}^C T^C\right) { \phi}-ig{\phi}^{\dagger}{\tilde A}_{\mu}^B T^BD^{\mu}\phi \nonumber \\
&&-ig\left(D^{\mu}{\tilde \phi}\right)^{\dagger}{\tilde A}_{\mu}^BT^B {\tilde \phi}+{ig}{\tilde \phi}^{\dagger}{\tilde A}_{\mu}^BT^B D^{\mu}{\tilde \phi}-g^2{\tilde \phi}^{\dagger}\left({\tilde A}_{\mu}^B T^B{\tilde A}_{\mu}^C T^C\right){\tilde \phi}. \nonumber
\end{eqnarray}
First we work in the higher derivative version of the theory. The gauge is fixed in the usual way introducing a parameter $\xi$ and the ${\hat A}$ propagator is,
\begin{equation}
{\hat D}^{AB}_{\mu \nu}(p)=-\delta^{AB}{i \over p^2-p^4/M_A^2}\left(\eta_{\mu \nu}-(1-\xi){p_{\mu}p_{\nu} \over p^2}-\xi {p_{\mu}p_{\nu} \over M_A^2}\right).
\end{equation}
We work in Landau gauge, $\xi=0$, which the propagator falls off as $1/p^4$ at large mometum $p$. Now a diagram that contributes to the ${\phi}$ mass has two external $\phi$ legs and it is straightforward to show, using Landau gauge in the higher derivative theory, that the degree of divergence of Feynman diagrams contributing to the ${\phi}$ mass is, $d=4-2L$, where $L$ is the number of loops.  At one loop, $L=1$, there is a possible quadratic divergence but not at two loops and beyond.  At one loop you can check by explicit calculation that there is no quadratic divergence. For this purpose it is easier to work in the formulation with LW fields and no higher derivative terms. In that version the propagator for the gauge bosons is,
\begin{equation}
D_{\mu \nu}^{AB}(p)=-\delta^{AB}{ i \over p^2}\left(\eta_{\mu \nu}-(1-\xi){p_{\mu}p_{\nu} \over p^2} \right),
\end{equation}
and the propagator for the massive LW vector in the adjoint representation is,
\begin{equation}
{\tilde D}_{\mu \nu}^{AB}(p)=\delta^{AB} {i \over p^2-M_A^2}\left(\eta_{\mu \nu}-{p_{\mu}p_{\mu} \over M_A^2}\right).
\end{equation}
There is a remaining logarithmic divergence and so corrections to the Higgs mass in the LW standard model have the form,
\begin{equation}
\delta m_h^2 \sim {g^2 M_{LW}^2 \over 16 \pi^2} {\rm ln}\left(\Lambda^2 \over M_{LW}^2\right),
\end{equation}
where $\Lambda$ is some ultraviolet cutoff where new physics beyond that in the LW extension of the standard model comes in (perhaps the Planck scale). One cannot have the LW partner masses $M_{LW}$ too much greater than a ${\rm TeV}$ and still solve the hierarchy puzzle.

 In the LW standard model there are no strong constraints on the mass matrices of the LW  particles from the smallness of flavor changing neutral currents~\cite{dulaney}.

The LW standard model is not finite but it is renormalizable and the one loop gauge coupling beta functions have been computed~\cite{og}.

\section{Constraints from Precision Electroweak Physics}

The magnitude of the LW masses for the $SU(2)\times U(1)$ $M_1$ and $M_2$ are constrained by precision electroweak physics~\cite{pe}. In the higher derivative version (neglecting the Higgs vacuum expectation value) of the LW standard model  the self energies of the weak gauge bosons take the form,
\begin{equation}
{\Pi}(q^2)=q^2-{q^4 \over M_{1,2}^2}.
\end{equation}
The precision electroweak parameters, $Y$ and $W$ are defined by,
\begin{equation}
Y={M_W^2 \over 2}\Pi_{BB}''(0),~~~~~W={M_W^2 \over 2}\Pi_{33}''(0).
\end{equation}
In the LW standard model at tree level,
\begin{equation}
Y=-{M_W^2 \over M_1^2}, ~~~~W={M_W^2 \over M_2^2}.
\end{equation}
Precision electroweak data imply, at the $99\%$ confidence level,  that $M_{1,2}> 3~{\rm TeV}$,  when we make the assumption that $M_1=M_2$.  Of course virtual LW particles  impact the values of other measurable quantities including properties of the Higgs boson~\cite{observe}.

\section{Fermions}

So far I have not mentioned how to put fermions in the LW version of the standard model.  Consider a theory of a massless left handed fermion  in the fundamental representation of the gauge group. The higher derivative theory in this case is,
\begin{equation}
\label{fermionhd}
{\cal L}_{\rm hd}={\bar {\hat \psi}_L}i \left(\gamma^{\mu} {\hat D}_{\mu}\right) {\hat \psi}_L+{1 \over M_{\psi}^2}{\bar {\hat \psi}_L}i \left(\gamma^{\mu} {\hat D}_{\mu}\right)\left(\gamma^{\nu}{\hat D}_{\nu}\right)\left(\gamma^{\eta} {\hat D}_{\eta}\right) {\hat \psi}_L.
\end{equation}
To eliminate the higher derivative terms we need in this case two auxillary  LW fields. One left handed (${\tilde \psi}_L$) and the other right handed (${\tilde \psi}_R$). With these LW-fields the fermion Lagrange density becomes,
\begin{eqnarray}
\label{lwfermion}
&&{\cal L}={\bar {\hat \psi}_L}i \left(\gamma^{\mu} {\hat D}_{\mu}\right) {\hat \psi}_L+M_{\psi}\left({\bar {\tilde \psi}_L}{\tilde \psi}_R+{\bar {\tilde \psi}_R}{\tilde \psi}_L\right)+{\bar {\tilde \psi}_L}i\left(\gamma^{\mu} {\hat D}_{\mu}\right){\hat \psi}_L \nonumber  \\
&&+{\bar {\hat \psi}_L}i\left(\gamma^{\mu} {\hat D}_{\mu}\right){\tilde \psi}_L-{\bar {\tilde \psi}_R}i\left(\gamma^{\mu} {\hat D}_{\mu}\right){\tilde \psi}_R.
\end{eqnarray}
Using the equations of motion,
\begin{equation}
{\tilde \psi}_R=- i{\gamma^{\mu}{\hat D}_{\mu} \over M_{\psi}}{\hat \psi}_L,~~~{\tilde \psi}_L={\gamma^{\mu}{\hat D}_{\mu} \gamma^{\nu}{\hat D}_{\nu}\over M_{\psi}^2}{\hat \psi}_L.
\end{equation}
Eq.~(\ref{lwfermion}) can be seen to be equivalent to Eq.~(\ref{fermionhd}). Finally, the kinetic terms can be put into standard form by performing the familiar shift,  $\hat \psi_L=\psi_L-{\tilde\psi}_L$. 

Of course to be consistent the LW fermions must develop a width and decay. This occurs because of the Yukawa couplings in the theory.  Some of the Yukawa's are very small, and hence some of the LW fermions are long lived (by particle physics standards). Because of this it may be possible to actually observe the acausal behavior discussed in section 2 in scattering experiments at the LHC~\cite{Alvarez:2009af}.

One popular method for giving neutrinos mass is the see-saw mechanism with heavy right handed neutrinos. It is possible to do this in the LW standard model and  solve the hierarchy problem with the right handed neutrinos  much heavier than the scale of LW particle masses~\cite{neut}.

\section{Conclusions}

LW- theories are quite interesting. They have a nonlocality in time or if you prefer an acausality that implies unusual behavior in scattering experiments. The order $p^4$ term in the propagators means these theories have softer behavior at high momentum than ordinary quantum field theories and therefore can help with the hierarchy problem and perhaps even with quantum gravity.  

If the LW standard model is the correct theory of nature you could eventually observe the unusual overall sign  and perhaps the unusual sign of the width term in the LW-gluon propagator from
interference between exchange of an ordinary gluon and LW-gluon in scattering amplitudes~\cite{rizzo}.   

In this talk I have considered LW theories with order $p^4$ terms in the propagators. It is possible to have even higher derivative terms\cite{carrone}. Also the fine tunning assocated with the Higgs mass can be reduced if only the particles that couple strongly to the Higgs have LW partners~\cite{little}. 

In constructing LW theories some of the usual rules of quantum field theory have been changed. The amplitudes have unusual analytic properties. It is certainly possible that further study will reveal that LW-theories are inconsistent. (As David Gross emphasized to me, ``the odds are not in your favor when you bet that a theory that changes the basic tenets of quantum field theory is consistent".) In my opinion the best chance for discovering a problem with LW theories is not in scattering amplitudes (which likely can be defined to all orders in perturbation theory to give a Lorentz invariant and unitary $S$-matrix) but in collective properties like the behavior of LW-theories at finite temperature.  

The LW toy model in section 2 has been studied~\cite{Fornal:2009xc} at finite temperature, $T$, by Fornal {\it et. al.}. They found that in the energy density and pressure at $T>>M$ the order $T^4$ term cancelled between the LW and ordinary scalar. This gives a speed of sound in the gas that approaches unity (from below) at very high $T$.  Currently, the authors of this paper are puzzling over the fermion case since naively there are twice as many LW fermion degrees of freedom (${\tilde \psi}_{L}$ and ${\tilde \psi}_R$)  as ordinary ones ($\psi_L$) and so the LW degrees of freedom overcompensate giving a negative energy density\footnote{I am gateful to Jose Ramon Espinosa for pointing this issue out.}. I hope this issue this will get resolved in the near future.

There are many more things to check and try to understand about LW-theories and the implications of the LW standard model for particle physics and for cosmology~\cite{cosmology}.  I certainly think they are interesting to study. 
%\footnote{However I am trying to restrict my work on this very speculative subject to roughly one paper per year. %Basically I am worried my colleagues will think I have turned into a ``nut case" if I work more on it.  I don't think its %crazy to work on these theories, but of course if I have turned into a ``nut case" I won't realize it.}. 
I hope this brief presentation will encourage others to think about LW theories and perhaps one day not too far in the future  more definitive statements can be made about their consistency as physical theories.

%%%%%%%%%%%%%%%%%%%%%%%%%%%%%%%%%%%%%%%%%%%%%%%%%%%%%%%%%%%%%%%%%%%%%%%%

\end{document}